\documentclass[twocolumn]{aastex631}

\newcommand*{\http}[1]{\href{http://#1}{#1}}

\newcommand{\drvm}{$\mathrm{\Delta RV_{max}\, } $}
\newcommand{\logg}{$\mathrm{log(g) } $}
\newcommand{\teff}{$\mathrm{T_{eff} } $}
\newcommand{\feh}{$\mathrm{[Fe/H] } $}
\newcommand{\aeh}{$\mathrm{[\alpha/H] } $}

\newcommand{\kms}{$\mathrm{km\,s^{-1}}$}

\graphicspath{{./}{figures/}}

\begin{document}

\title{Multiplicity Statistics of Stars in the Sagittarius Dwarf Spheroidal Galaxy: Comparison to the Milky Way}

\author[0000-0001-5330-7709]{Victoria Bonidie}
\affiliation{Department of Physics and Astronomy, University of Pittsburgh, 3941 O'Hara Street, Pittsburgh, PA 15260, USA}
\affiliation{Pittsburgh Particle Physics, Astrophysics, and Cosmology Center (PITT PACC), University of Pittsburgh, Pittsburgh, PA 15260, USA}

\author[0000-0003-3837-7201]{Travis Court}
\affiliation{Department of Physics and Astronomy, University of Pittsburgh, 3941 O'Hara Street, Pittsburgh, PA 15260, USA}
\affiliation{Pittsburgh Particle Physics, Astrophysics, and Cosmology Center (PITT PACC), University of Pittsburgh, Pittsburgh, PA 15260, USA}

\author[0000-0003-2116-2159]{Christine Mazzola Daher}
\affiliation{Department of Physics and Astronomy, University of Pittsburgh, 3941 O'Hara Street, Pittsburgh, PA 15260, USA}
\affiliation{Pittsburgh Particle Physics, Astrophysics, and Cosmology Center (PITT PACC), University of Pittsburgh, Pittsburgh, PA 15260, USA}

\author[0000-0001-8245-779X]{Catherine E. Fielder}
\affiliation{Department of Physics and Astronomy, University of Pittsburgh, 3941 O'Hara Street, Pittsburgh, PA 15260, USA}
\affiliation{Pittsburgh Particle Physics, Astrophysics, and Cosmology Center (PITT PACC), University of Pittsburgh, Pittsburgh, PA 15260, USA}

\author[0000-0003-3494-343X]{Carles Badenes}
\affiliation{Department of Physics and Astronomy, University of Pittsburgh, 3941 O'Hara Street, Pittsburgh, PA 15260, USA}
\affiliation{Pittsburgh Particle Physics, Astrophysics, and Cosmology Center (PITT PACC), University of Pittsburgh, Pittsburgh, PA 15260, USA}

\author[0000-0001-8684-2222]{Jeffrey Newman}
\affiliation{Department of Physics and Astronomy, University of Pittsburgh, 3941 O'Hara Street, Pittsburgh, PA 15260, USA}
\affiliation{Pittsburgh Particle Physics, Astrophysics, and Cosmology Center (PITT PACC), University of Pittsburgh, Pittsburgh, PA 15260, USA}

\author[0000-0002-0870-6388]{Maxwell Moe}
\affiliation{Steward Observatory, University of Arizona, 933 N. Cherry Avenue, Tucson, AZ 85721, USA}

\author[0000-0001-5253-1338]{Kaitlin M. Kratter}
\affiliation{Steward Observatory, University of Arizona, 933 N. Cherry Avenue, Tucson, AZ 85721, USA}

\author[0000-0003-2496-1925]{Matthew G. Walker}
\affiliation{McWilliams Center for Cosmology, Department of Physics, Carnegie Mellon University, 5000 Forbes Avenue, Pittsburgh, PA 15213, USA}

\author[0000-0003-2025-3147]{Steven R. Majewski}
\affiliation{Dept. of Astronomy, University of Virginia, P.O. Box 400325, Charlottesville, VA 22904-4325 USA}

\author[0000-0003-2969-2445]{Christian R. Hayes}
\affiliation{NRC Herzberg Astronomy and Astrophysics, 5071 West Saanich Road, Victoria, B.C., Canada, V9E 2E7}

\author[0000-0001-5388-0994]{Sten Hasselquist}
\affiliation{New Mexico State University, Las Cruces, NM 88003, USA}

\author[0000-0002-3481-9052]{Keivan Stassun}
\affiliation{Department of Physics and Astronomy, Vanderbilt University, Nashville, TN 37235, USA}

\author[0000-0002-5365-1267]{Marina Kounkel}
\affiliation{Department of Physics and Astronomy, Vanderbilt University, Nashville, TN 37235, USA}

\author[0000-0001-6977-9495]{Don Dixon}
\affiliation{Department of Physics and Astronomy, Vanderbilt University, Nashville, TN 37235, USA}

\author[0000-0003-1479-3059]{Guy S. Stringfellow}
\affiliation{ Center for Astrophysics and Space Astronomy, Department
of Astrophysical and Planetary Sciences, University of Colorado,
389 UCB, Boulder, CO 80309-0389, USA}

\author[0000-0001-5926-4471]{Joleen K. Carlberg}
\affiliation{Space Telescope Science Institute, 3700 San Martin Drive, Baltimore, MD 21218, USA}

\author[0000-0001-5261-4336]{Borja Anguiano}
\affiliation{Department of Astronomy, University of Virginia, Charlottesville, VA, 22904, USA}

\author[0000-0002-3657-0705]{Nathan De Lee}
\affiliation{Department of Physics, Geology, and Engineering Tech, Northern Kentucky University, Highland Heights, KY 41099, USA}

\author[0000-0003-3248-3097]{Nicholas W. Troup}
\affiliation{Department of Physics, Salisbury University, Salisbury, MD 21801, USA}

\begin{abstract}
We use time-resolved spectra from the Apache Point Observatory Galactic Evolution Experiment (APOGEE) to examine the distribution of radial velocity (RV) variations in 249 stars identified as members of the Sagittarius (Sgr) dwarf spheroidal (dSph) galaxy by \cite{Hayes}.  We select Milky Way (MW) stars that have stellar parameters (\logg, \teff, and \feh) similar to those of the Sagittarius members by means of a k-d tree of dimension 3. We find that the shape of the distribution of RV shifts in Sgr dSph stars is similar to that measured in their MW analogs, but the total fraction of RV variable stars in the Sgr dSph is larger by a factor of $\sim 2$. After ruling out other explanations for this difference, we conclude that the fraction of close binaries in the Sgr dSph is intrinsically higher than in the MW. We discuss the implications of this result for the physical processes leading to the formation of close binaries in dwarf spheroidal and spiral galaxies.

%\tori{which is that multiplicity statistics of stars formed in the Sgr dSph follow the same trend of stars formed in a typical disk galaxy like the MW, but are systematically above the median values of the MW $\mathrm{\Delta RV_{max}}$ values. But our sample is small so we cannot make any conclusions about their binary formation rate? also the difference in alpha abundances may be contributing significantly to this? help}

\end{abstract}

\section{Introduction} \label{sec:intro}
The Sagittarius (Sgr) dwarf spheroidal (dSph) galaxy \citep{Ibata} is one of the closest and largest satellites of the Milky Way (MW), and presents a unique opportunity to explore the stellar content of a galaxy outside of our own in detail. Recent studies have used a combination of 3D positions and kinematics to identify nearby stars as members of the Sgr dSph \citep{Has, Hayes}. Many of these objects now have multi-epoch, high signal-to-noise, high-resolution spectroscopy from the Apache Point Observatory Galactic Evolution Experiment (APOGEE; \citealt{Majewski2017, Wilson2019}), one of the core surveys of the Sloan Digital Sky Survey IV (SDSS-IV; \citealt{Gunn2006,Blanton2017}). Most of the Sgr dSph members observed by APOGEE also have high quality parallaxes from Gaia \citep{gaia, sanders+das+18}, and are thus among the best-studied stars known to have formed outside our Galaxy.  

Here we leverage the opportunity offered by these observations to examine the stellar multiplicity statistics in the Sgr dSph. The multi-epoch feature of the APOGEE spectra allows us to measure high precision radial velocities (RVs; \citealt{Nidever2015}) from Doppler shifts in the visit spectra, and define a maximum RV shift, $\mathrm{\Delta RV_{max} = | RV_{max} - RV_{min} |}$, for each star. This figure of merit is a reliable indicator for the presence of short-period binary companions, and it can be used to put constraints on multiplicity statistics \citep[see][for discussions]{B+M2012, M+B+B2012}. Previous studies have shown that RV variability is a strong function of stellar parameters \citep{moe+di+2017}. For the cool (\teff$\lesssim 6500$ K) stars targeted by APOGEE, the parameters with the strongest correlation to RV variability are \logg\ (due to the attrition of the shortest period companions as stars ascend the Red Giant Branch, \citealt{Badenes2018}) and chemical composition (due to the strong anti-correlation between the close binary fraction, metallicity and $\alpha$ abundances, \citealt{Badenes2018, Moe+19, Mazzola}), with \teff\ playing a more secondary role \citep{Mazzola, Price-Whelan+20}. 

A complete characterization of the relationship between stellar parameters, \drvm distributions, and multiplicity statistics requires large sample sizes for a robust multivariate analysis \citep[e.g.][]{Mazzola}, but this approach is not feasible for the small number of stars identified as members of the Sgr dSph. Here we propose an alternative method to constrain multiplicity statistics using relatively small samples. This method takes advantage of the large number of MW stars observed with APOGEE to define a control sample of stellar analogs matched on physically-motivated parameters. A statistical comparison between the RV shifts in the sample under study and the MW analogs can reveal differences in the underlying multiplicity statistics, which, if present, should be caused by parameters that have not been controlled for in the generation of the analog sample.

The characterization of the fundamental statistics of stellar multiplicity (the distribution of periods, eccentricities, and mass ratios) remains an important problem in stellar astrophysics \citep{moe+di+2017}. It is a particularly relevant issue for dSph galaxies, given that inferences on the dark matter content of these objects can be easily biased by the presence of a few short-period binaries \citep{mcconnahie+cote+10, koposov+11, Buttry+21}. Multi-epoch RV data sets have been used in the past to put constraints on the multiplicity statistics of dSph galaxies \citep[e.g.][]{martinez+11, simon+11}, but our goal is to make a direct comparison of the RV variability statistics between two samples of similar stars in the Sgr dSph and the MW. This comparison will help us elucidate whether the physical mechanisms responsible for the formation of close binaries operate differently in dark-matter dominated dwarf galaxies and star-forming spirals \citep[e.g.][]{minor+19, wyse+20}. 
%\redpen{These will do for now. Maybe some of our co-authors can suggest better ones.} 

%The Apache Point Observatory Galactic Evolution Experiment 2 (APOGEE-2; \citealt{Majewski2017}) survey within the Sloan Digital Sky Survey IV (SDSS-IV; \citealt{Gunn2006,Blanton2017}) uses the Doppler shifts in detected in the spectra to measure high precision radial velocities (RVs; \citealt{Nidever2015}). 
%\cmd{Don Schneider has really been getting after people to cite Wilson et al. 2019, the latest paper on the spectrograph (righfully, John Wilson + co deserve tons of credit). Can we work this in some way, or somewhere else? I have a feeling we'll get flagged if we don't cite it}

This paper is organized as follows. In section \ref{subsection: Sagittarius Sample Selection}, we
describe the sample of Sgr dSph members from \citet{Hayes} and the quality cuts we apply to measure RV variability. In section \ref{subsection: Analog Sample Selection}, we discuss the method used to select the MW analog sample, which we compare to the Sgr dSph sample in section \ref{results}, along with a brief discussion of the physical interpretation of this comparison.

%we compare these two samples -- the Sgr stars and their analogs in the MW -- via robust statistics to confirm that they do in fact mirror each other on those parameters and are not being influenced by some phenomenon that could cause systematic biases. After comparing the two samples on the parameters upon which we select analogs, we compare their $\mathrm{\Delta RV_{max}}$ distributions to see if the Sgr dSph stars and the MW analogs are statistically similar on that parameter. Finally, in section \ref{section: Results} we present our argument that the multiplicity statistics of stars formed in the Sgr dSph are similar to those of stars formed in a typical disk galaxy like the MW, but are systematically above the median values of the MW $\mathrm{\Delta RV_{max}}$ values.

\section{Methods}
\label{section: Methods}
\subsection{Sagittarius Sample Selection}
\label{subsection: Sagittarius Sample Selection}

The sample of Sgr dSph members published by \cite{Hayes} relies on APOGEE data release 16 \citep{Ahumada2020, Joensson2020}, and adds 518 new stars to the 325 previously identified in \citet{Has}, for a total of 876 members. These stars were selected using a combination of chemical tagging and 3D positions and velocities relative to the Sgr orbital plane, with additional cuts imposed to minimize contamination from MW halo stars. The identified Sgr members sample both the core of the dSph galaxy and the trailing and leading arms of the stream.

%as a selection criterion for their sample. Because we expect Sgr stars to have conserved their angular momentum, they should not have large velocities perpendicular to the Sgr orbital plane. Thus, the strictest selection parameters applied by \citet{Hayes} are a cut in the angular momentum of stars in our sample along the z-direction to remove MW contamination, and a cut in velocities to remove stars with high velocities perpendicular to the Sgr orbital plane. The details of this selection are outlined in \cite{Hayes}.

% In addition, \citet{Hayes} only wants to include stars with low velocity uncertainty, so they apply a cut at $\mathrm{V_{err} \leqslant 0.2  km/s}$. They are left with 518 new members of the Sgr system, including 133 new stars from the stream and 385 new stars from the core (referred to as Sgr dSph core stars). The sample is combined with 325 previously determined Sgr dSph stars from \cite{Has} to give a total sample of 876 APOGEE  stars in the Sgr system. We apply these same quality cuts to the full DR16 sample, which includes both stars in the Sgr dSPh and in the Milky Way.

In order to make robust measurements of RV variability for these Sgr dSph members, we introduced the quality cuts described in \citet{Mazzola}, which remove stars targeted as telluric calibrators, star cluster members, and commissioning stars, and require acceptable ($\neq 9999$) values for \teff, \logg, and \feh\ in the APOGEE Stellar Parameter and Chemical Abundances Pipeline \citep[ASPCAP][]{Perez2016,Joensson2020}. For the RV measurements, we used the \textsc{visits\_pk} indices \citep{Holtzman2015,Nidever2015} to select the individual visits that were included in each combined APOGEE spectrum, and we required that each star have two or more visit spectra with S/N $\geq 40$. A total of 249 Sgr dSph members survived these cuts, 48 in the extended stellar stream and 201 in the core \citep{Has}.

\subsection{Analog Sample Selection}
\label{subsection: Analog Sample Selection}

To select stellar analogs to the Sgr dSph members, we begin by imposing the combined quality cuts from \cite{Hayes} and \cite{Mazzola} on all the stars in APOGEE DR16. This results in an initial sample of 188,104 MW stars for which we can extract stellar parameters and RV measurements of at least the same quality as the Sgr dSph members. Since most MW stars are closer and brighter than the Sgr dSph members, their spectra can have much larger signal-to-noise ratios (SNR). To ensure that the presence of MW stars with very high SNR spectra does not introduce unwanted biases in our measurements, we require that the SNR of the stellar analogs be in the same range as the Sgr dSph members (53$<$SNR$<$283), leaving us with a sample of 139,983 MW stars from which to draw our analogs. We note that this additional restriction on the MW analogs does not affect the results described below.

%We run the same analysis using the both the full MW sample and the SNR-cut sample and obtained consistent results, allowing us to conclude that this restriction does not affect the results described below. We focus on the SNR-cut sample for the rest of this discussion, as this sample help to mitigate the concern of RV uncertainties being systematically and substantially different between the samples. \tori{Is this enough information on the non-cut sample?}

\begin{figure*}
    \centering
    \includegraphics[width=0.8\textwidth]{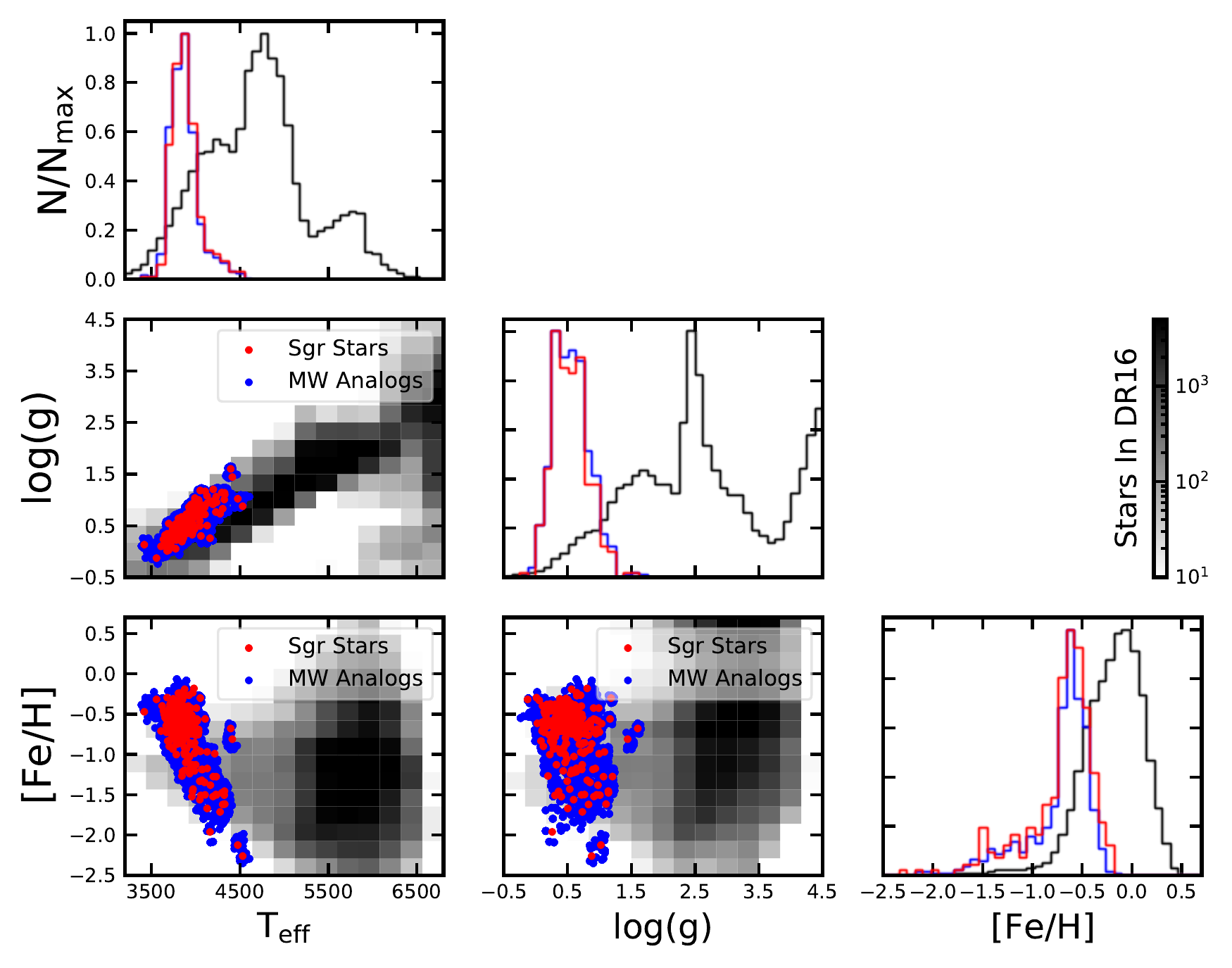}
    \caption{Triangle plot of the pair-wise distributions of \teff, \logg, and \feh\ for the full APOGEE DR16 sample (shaded two-dimensional histograms), along with the Sgr dSph members (red points) and their MW analogs (blue points). The one-dimensional histograms along the diagonal show the normalized distributions for each parameter individually. 
    %\tori{does this plot load faster now? and does it look too big?}
    }
    \label{triangle plot}
\end{figure*} 

%The statistical method for selecting analogs was presented first in \citet{licquia2015}. In a nutshell,

Our method to select stellar analogs to the Sgr dSph members from the main APOGEE DR16 sample is based on methods described in \cite{licquia2015} and \cite{fielder2021}. The goal is to define a large enough sample of MW stars whose parameters have statistical distributions that are similar to those of the Sgr dSph members. Because RV variability in APOGEE targets is driven mainly by location in the Hertzsprung-Russell diagram and chemical composition \citep{Mazzola}, we choose \logg, \teff, and \feh\ as the parameters for analog selection. We note that the relationship between stellar multiplicity and chemical composition is complex, and that $\alpha$ abundances in particular can have a strong impact on multiplicity statistics at fixed \feh\ \citep[see][for a discussion]{Mazzola}. However, we find that the introduction of additional selection parameters results in non-viable analog sample sizes. We will revisit the impact of $\alpha$ abundances on our measurements in Section~\ref{results}.

To optimize the search for analogs in multidimensional space, we use a k-d tree, which is a binary tree with points in k-dimensional space \citep{bentley1975}. Given a star A in the Sgr dSph member sample, we standardize the values for all the stars in our main MW sample, subtracting from each parameter the value measured in star A and dividing by its standard deviation (i.e., the measurement uncertainty). This ensures that all data are on the same scale when put into the tree. We then use the \texttt{spatial.cKdDTree} routine from \citet{SciPy2020} to construct a k-d tree from this scaled data, which we query to find the closest $N$ neighbors to star A in the three-dimensional space of selection parameters. 
%We perform a Kolmogorov–Smirnov (K-S) statistical test for \teff, \logg, and \feh\ between samples as a function of neighbor number and find that choosing 20 neighbors had the optimum convergence. 
We performed a series of a Kolmogorov–Smirnov (K-S) tests on the distributions of \teff, \logg, and \feh\ between the analog and Sgr samples as function of neighbor number, and we found that $N=$20 offers a good compromise between sample size and similarity of the resulting distributions. To further restrict the distance of the selected neighbors,
%we impose an upper bound to the distance of the selected neighbors, which must be 
we require the analog candidate's parameters to fall within one standard deviation of the mean parameters of star A.

After the 20 nearest neighbors to star A are selected, a single analog must be chosen among them. To do this, we assign weights to each neighbor based on their distance in parameter space from the query point. Our weights ($w$) are equal to 
\begin{equation}
w = \frac{e^{-d^{2}/2}}{\sum({e^{-d^{2}/2}})},
\end{equation}
where $d$ is the Euclidean distance in parameter space to the neighbor (e.g., $d(\vec{x},\vec{y}) = \sqrt{\sum_{i=1}^{n}(x_{i}-y_{i})^{2}}$ for two n-dimensional vectors). The analog to star A is then selected from a random draw amongst these neighbors, using these weights. The selection then proceeds as outlined in \citet{licquia2015}. Using this procedure, we obtain 100 MW analogs for each star in the Sgr dSph member sample, resulting in a total sample of 24,900 analog stars, of which 1,679 are unique. Repetition of analogs for each star is necessary to ensure that the probability distributions of the selection parameters in the analog sample closely match those in the Sgr dSph sample \citep{licquia2015}. 
%\cef{This is a statistical effect, duplicates are necessary in order to ensure that the distribution of properties of the analogs matches with the posterior distribution of each star's trio of properties. I think you can cite \citet{licquia2015} for this and if you'd like to, this method is inspired by the "sosies" method of Bottinelli et al 1995 and de Vaucouleurs \& Corwin 1986.}. 
The Galactic analog search code is publicly available at the \href{https://github.com/cfielder/Milky-Way-Analogs}{Milky Way analog GitHub page}.\footnote{https://github.com/cfielder/Milky-Way-Analogs. This code is provided under a CC BY-SA 4.0 license} The stellar version of the code, used in the work presented here, is available upon request.

Figure \ref{triangle plot} shows a comparison between the stars in the Sgr dSph member sample and our MW analog sample in the three-dimensional space of selection parameters, along with the main APOGEE DR16 sample, which is included for illustrative purposes. As expected, the distribution of values in the MW analog sample closely matches that of the Sgr dSph sample on all three selection parameters ($\mathrm{T_{eff}}$, $\mathrm{log(g)}$, and $\mathrm{[Fe/H]}$). A K-S test on the selection parameter distributions for the two samples yields p-values of 0.496 for $\mathrm{T_{eff}}$, 0.756 for $\mathrm{log(g)}$, and 0.072  for $\mathrm{[Fe/H]}$. Since all p-values are above  $0.05$, we conclude that the samples are well matched. Our MW analog sample spans the entire range of RA and Dec in DR16, indicating that we are not preferentially selecting stars from any specific regions or substructures in the Milky Way. We also confirm that the distributions of the number of visits are similar between both samples.

\section{Results and Discussion: Stellar Multiplicity in the Sagittarius Dwarf Spheroidal}\label{results}

\begin{figure}
    \centering
\includegraphics[width=\columnwidth]{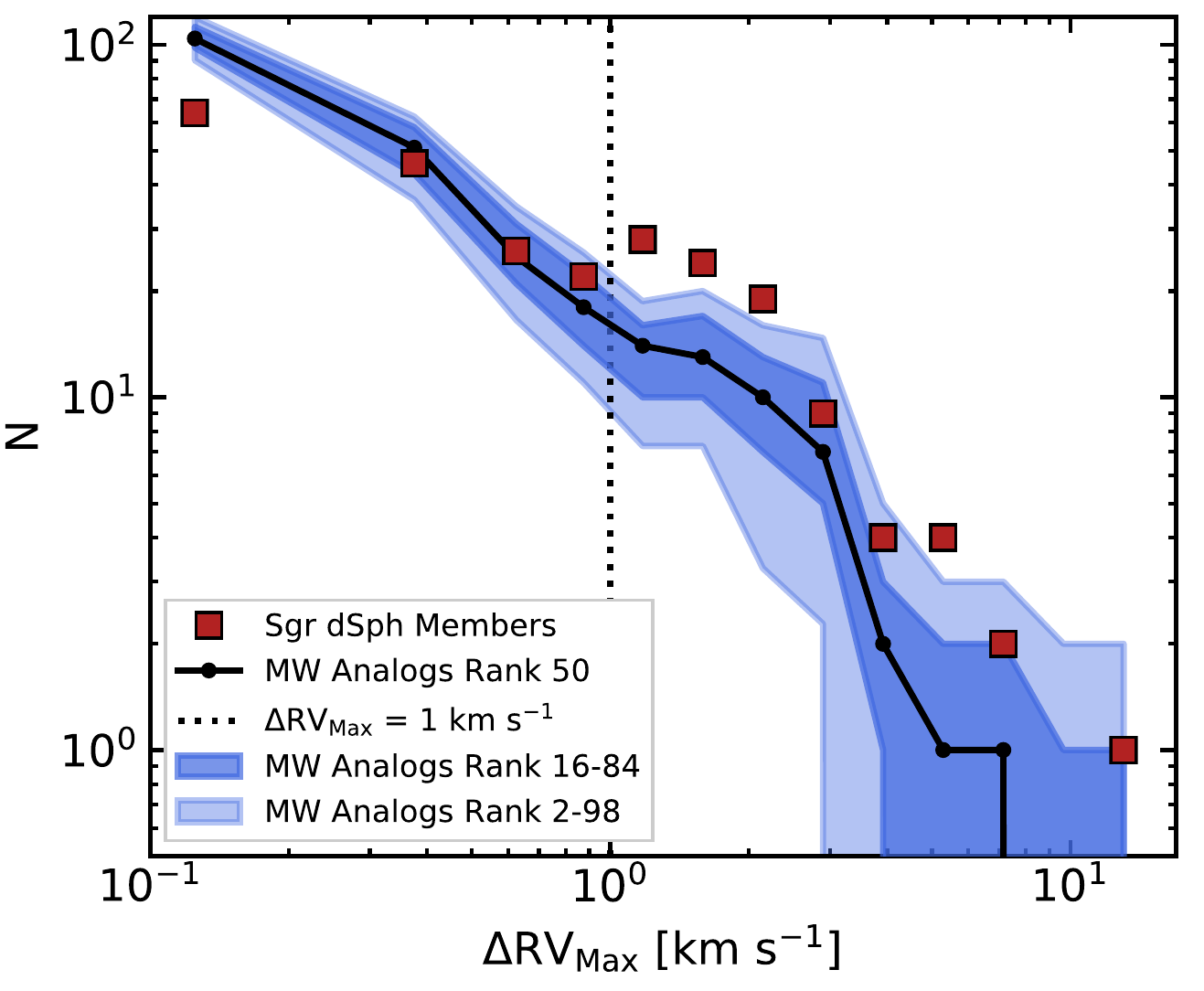}
    \caption{Histogram of \drvm\ values measured in the Sgr dSph members (red squares), together with the range of values found in the MW analogs (solid black line for the 50th rank, dark and light shaded regions for the 16th-84th and 2nd-98th ranks, respectively). For illustrative purposes, we highlight a \drvm\ value of 1 \kms\ with a vertical dotted line.}
    \label{norm_hist}
\end{figure}

We now have two samples of metal-poor K-type giants that are closely matched on three selection parameters (\logg, \teff, and \feh), one composed of Sgr dSph members and one of MW stars. We can compare the multiplicity statistics of these two samples through the distributions of \drvm\ values, which are shown in Figure \ref{norm_hist}. In this plot, the red squares represent the histogram of \drvm\ values in the Sgr dSph sample, while the shaded region represents the statistical range (50th rank in black, 16th-84th ranks in dark blue, and 2nd-98th ranks in light blue) of the \drvm\ distributions found in the 100 MW analog sub-samples with $N=N_{Sgr}=249$ that can be constructed by random draw from the main MW analog sample.

\begin{figure}
    \centering
    \includegraphics[width=\columnwidth]{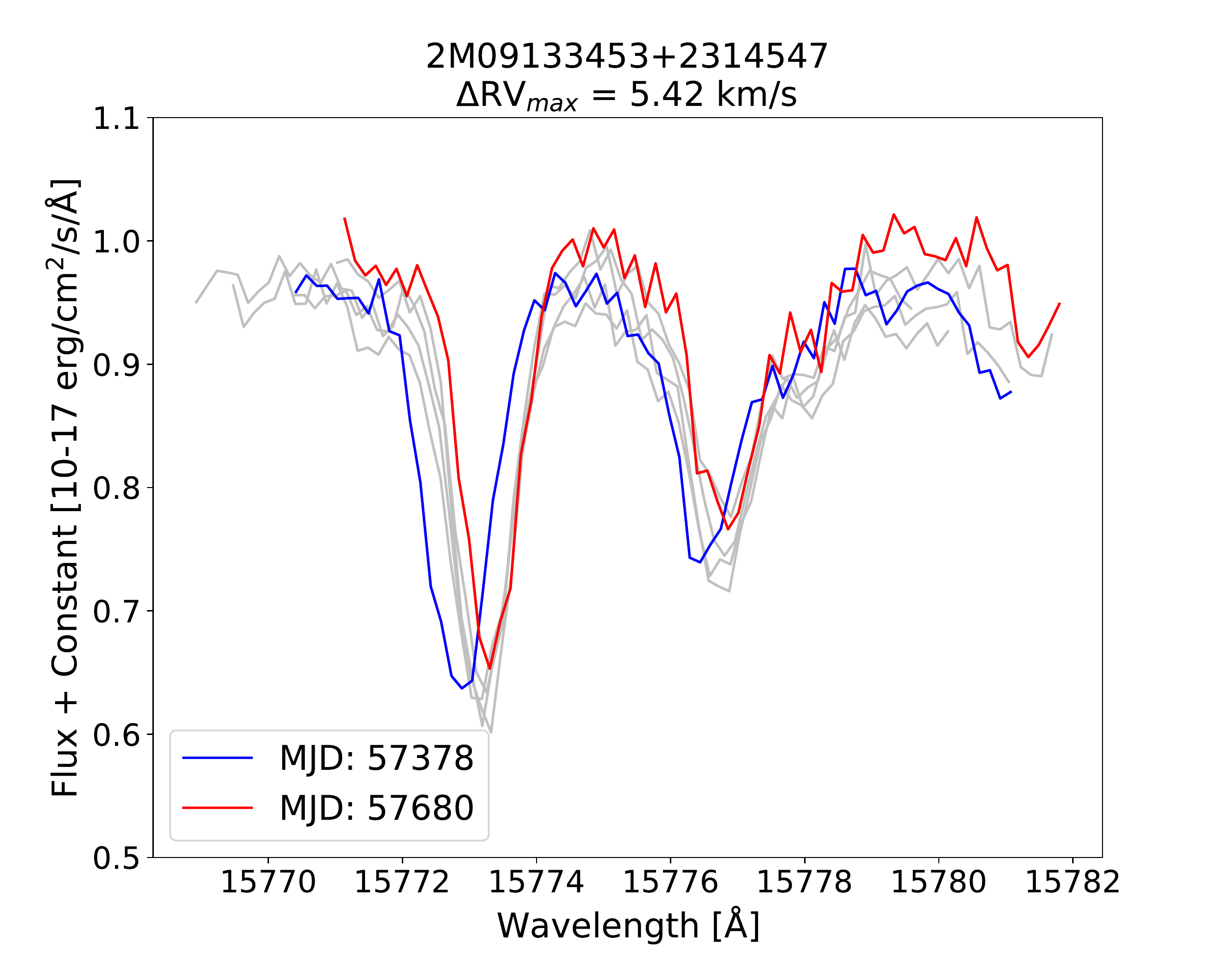}
    \caption{The visit spectra for the Sgr dSph member, 2M09133453+2314547, shown around the Fe II line at 15,773.5860 {\AA} and the Fe I line at 15,776.733 {\AA} \citep{NIST_ASD}. The \drvm\ for this star is 5.42 \kms, corresponding to the difference between the visit spectra highlighted in red (MJD 57680) and blue (MJD 57378).}
    \label{spectra}
\end{figure} 

The distribution of \drvm values in a sample of stars with multiple RV measurements of good quality is characterized by a core of systems with low \drvm, composed of wide binaries and single stars, and a distinct tail of systems at high \drvm\ that is dominated by short-period binaries. The shape and extent of the core, and the location of the core-tail transition, are mainly driven by the distribution of RV errors, which can be difficult to characterize (see \citealt{M+B+B2012} and \citealt{Mazzola} for discussions). \cite{Badenes2018} found that the dwarfs and most of the giants observed by APOGEE have core/tail transitions around 1 or 2 \kms, but metal-poor giants close to the tip of the RGB can have broader cores (see their Figure 9 and related discussion in their section 3.3). However, as long as both the Sgr dSph member sample and the MW analog sample contain enough \textit{bona fide} short-period binaries at high values of \drvm, it is not necessary to specify a threshold value to compare the distributions. To ensure that this is the case, we visually inspected the visit spectra for each Sgr dSph member with \drvm $>$ 3 \kms, and confirmed they all have clear Doppler shifts; we show an example in Figure \ref{spectra} with $\mathrm{\Delta RV_{max}}$ = $5.42$ \kms. 
%This is just one of a handful of low-mass spectroscopic potential binaries believed to have formed outside the MW \citep[e.g.][]{Buttry+21, Lewis+20, koch+14}. 
Unfortunately, none of our Sgr stars appear in the \textit{Gold} sample from \cite{Price-Whelan+20}, so we do not have any estimates for their orbital parameters from \textit{The Joker} \cite{Price-Whelan2017}. This is unsurprising, given that the vast majority of the stars in our Sgr sample have very sparsely-sampled RV curves (76\% have only 2 or 3 RVs). Regardless, the presence of \textit{bona fide} short-period binaries like the system shown in Fig.~\ref{spectra} confirms that the \drvm\ distributions can be used to constrain the multiplicity statistics of the underlying stellar populations.

%This means that the \drvm\ distributions can be used to put constraints on the underlying multiplicity statistics.

%The presence of short-period binaries like this one confirms that the tails of the \drvm\ distributions shown in Figure \ref{norm_hist} can be used to constrain the multiplicity statistics of the underlying stellar populations.

\begin{figure}
    \centering
    \includegraphics[width=\columnwidth,trim = 25 15 80 63, clip]{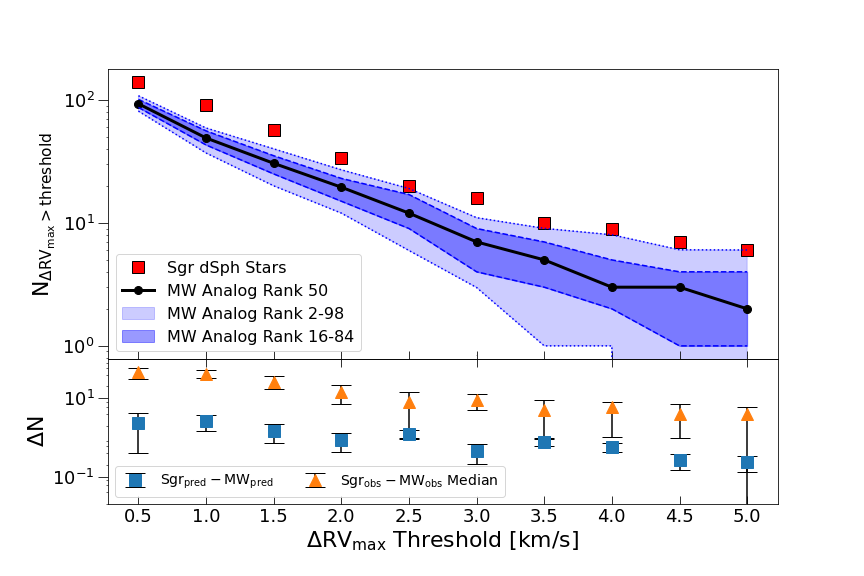}
    \caption{Upper panel: Number of Sgr dSph stars with $\mathrm{\Delta RV_{max}}$ above the given threshold (red squares), together with the 50th rank (black line), 16th-84th ranks (dark blue region) and 2nd-98th ranks (light blue region) of values seen in the MW analog samples. Lower panel: Difference between the number of RV variable stars measured in the Sgr and MW analog samples (orange triangles), alongside our estimate for the difference that would arise from the offset of \aeh\ values between the samples (blue squares).}
    \label{drvm_thresh}
\end{figure}

A direct comparison between the tails of the \drvm\ distributions of the Sgr dSph stars and their MW analogs by means of an Anderson-Darling test yields a p-value below 0.001, which is the floor of the test. This allows us to reject the null hypothesis that the underlying distributions are identical. Indeed, even though the high \drvm\ tails shown in Figure \ref{norm_hist} have similar shapes, the Sgr dSph stars clearly have a larger RV variability fraction, with star counts always above the median and often beyond the 98th rank of the MW analogs. We further illustrate this point in Figure \ref{drvm_thresh}, which shows the total number of stars with \drvm\ above a certain threshold, as a function of the threshold value. The number of RV variable stars in the Sgr dSph sample is consistently above the 98th rank of the numbers seen in the MW analog samples. Crucially, this does not depend on the value of the threshold we choose for RV variability (i.e., the location of the vertical dotted line in Figure~\ref{norm_hist}), well into the regime where we have visually inspected every star to confirm that the RV shifts are real. This means that an extended tail of RV errors cannot be responsible for the higher fraction of RV variables seen in the Sgr dSph sample. we also note that the ratio of Sgr dSph to MW analog RV variables does not change when the samples are split into high and low \feh\ along the median.

To explain the larger fraction of RV variables among Sgr dSph stars without invoking different multiplicity statistics, the RV variability would have to be linked to a parameter or parameters that were not controlled for in the analog selection process. Among the parameters that were left out of the k-d tree, $\alpha$ abundances have the strongest correlation with RV variability in APOGEE targets \citep{Mazzola}. Indeed, the values of \aeh\ in the Sgr dSph stars are $\sim 0.13$ dex lower than those of their MW analogs, which is consistent with the known chemistry of the Sgr dSph galaxy \citep{Sbordone07}, and should lead to higher RV variability \citep{Mazzola}. To quantify the impact that this offset in [$\alpha$/H] might have on RV variability, we used a sample of 31,176 giants in APOGEE (\logg\ $< 1.5$, \teff\ $< 4500$ K) to estimate the RV variability fractions of the Sgr and MW analog samples based on their \feh\ and \aeh\ values. For each Sgr and MW analog star, we identified the bin in \feh-\aeh\ space that it corresponds to in the giants sample, and assigned it the RV variability fraction and associated uncertainty of that bin. These weighted averages can be used to estimate the increase in RV variability that we would see just from the fact that the \aeh\ values of stars in the Sgr sample are lower than those in their MW analogs. This increase is shown in the lower panel of Figure~\ref{drvm_thresh}, and is in all cases smaller than the actual difference in RV variability measured between the Sgr and MW analog samples.

We conclude that the increased fraction of RV variables seen among Sgr dSph stars is too large to be explained by the systematic difference in \aeh\ between those stars and their MW analogs. Our analysis suggests that the close binary fraction in the Sgr dSph sample is in fact higher than in the MW analogs, by a factor $\sim$2, which is relatively insensitive to the choice of \drvm\ threshold. The physical reason for this higher fraction of close binaries is unclear at this stage. \citet{wyse+20} found an enhanced occurrence rate of blue stragglers in metal-poor dSph galaxies, including Sgr, which they attributed to the metallicity-dependence of the close binary fraction and the outcome of processes such as dynamical friction and mass segregation. Mass segregation through two-body interactions, which drives heavier binaries to the bottom of the potential well, is known to operate in relaxed stellar systems like globular clusters and ultrafaint dwarf galaxies, that are not dynamically dominated by dark matter \citep{Wu2021,Baumgardt22}, driving the heavier binaries to the bottom of the potential well. However, given the properties of the Sgr dSph \citep{Gibbons2017}, it is unlikely that the system was relaxed before being accreted by the MW \citep{2008gady.book.....B}. In any case, the number of stars from the Sgr dSph stream in our sample (48) is too small to firmly establish whether the fraction of RV variables among them is higher or lower than in the core of the Sgr dSph, as would be predicted by this scenario. In our sample, the fraction of stars with \drvm $> 1$ km/s in the stream is $0.40 \pm 0.03$ and that of the core is $0.36 \pm 0.07$. The error bars on the RV variability fractions are too large, and their behavior as a function of \drvm\ is too noisy, to confirm or rule out mass segregation as the origin of the higher RV variability in the Sgr dSph. Finally, we note that the methods we have described in this paper can be applied to other systems in the Local Group with similar data from APOGEE and other samples, like the Magellanic Clouds or the Gaia-Enceladus sausage \citep{sanders+21}. In this way, it would be possible to explore systematic variations in stellar multiplicity statistics across different dynamical environments using relatively small samples of sparsely sampled RV curves.

\section*{Acknowledgements}

VB acknowledges support from a Whittington Fellowship awarded by the Dietrich School of Arts and Sciences at the University of Pittsburgh. CMD and CB acknowledge support from the National Science Foundation grant AST-1909022. CEF acknowledges support from NASA Astrophysics Data Analysis Program grant number 80NSSC19K0588.

This work made use of Python, along with many community-developed or maintained software packages, including IPython \citep{ipython}, Jupyter (\http{jupyter.org}), Matplotlib \citep{matplotlib}, NumPy \citep{numpy}, Pandas \citep{pandas}, scikit-learn \citep{scikit-learn}, and SciPy \citep{SciPy2020}.
This research made use of NASA's Astrophysics Data System for bibliographic information.

\bibliography{bibliography}{}

\begin{thebibliography}{}
\expandafter\ifx\csname natexlab\endcsname\relax\def\natexlab#1{#1}\fi
\providecommand{\url}[1]{\href{#1}{#1}}
\providecommand{\dodoi}[1]{doi:~\href{http://doi.org/#1}{\nolinkurl{#1}}}
\providecommand{\doeprint}[1]{\href{http://ascl.net/#1}{\nolinkurl{http://ascl.net/#1}}}
\providecommand{\doarXiv}[1]{\href{https://arxiv.org/abs/#1}{\nolinkurl{https://arxiv.org/abs/#1}}}

\bibitem[{Ahumada {et~al.}(2020)Ahumada, Prieto, Almeida, Anders, Anderson,
  Andrews, Anguiano, Arcodia, Armengaud, Aubert, Avila, Avila-Reese, Badenes,
  Balland, Barger, Barrera-Ballesteros, Basu, Bautista, Beaton, Beers,
  Benavides, Bender, Bernardi, Bershady, Beutler, Bidin, Bird, Bizyaev, Blanc,
  Blanton, Boquien, Borissova, Bovy, Brandt, Brinkmann, Brownstein, Bundy,
  Bureau, Burgasser, Burtin, Cano-D{\'\i}az, Capasso, Cappellari, Carrera,
  Chabanier, Chaplin, Chapman, Cherinka, Chiappini, Doohyun~Choi, Chojnowski,
  Chung, Clerc, Coffey, Comerford, Comparat, da~Costa, Cousinou, Covey, Crane,
  Cunha, Ilha, Dai, Damsted, Darling, Davidson, Davies, Dawson, De, de~la
  Macorra, De~Lee, Queiroz, Deconto~Machado, de~la Torre, Dell'Agli, du~Mas~des
  Bourboux, Diamond-Stanic, Dillon, Donor, Drory, Duckworth, Dwelly, Ebelke,
  Eftekharzadeh, Davis~Eigenbrot, Elsworth, Eracleous, Erfanianfar, Escoffier,
  Fan, Farr, Fern{\'a}ndez-Trincado, Feuillet, Finoguenov, Fofie,
  Fraser-McKelvie, Frinchaboy, Fromenteau, Fu, Galbany, Garcia,
  Garc{\'\i}a-Hern{\'a}ndez, Oehmichen, Ge, Maia, Geisler, Gelfand, Goddy,
  Gonzalez-Perez, Grabowski, Green, Grier, Guo, Guy, Harding, Hasselquist,
  Hawken, Hayes, Hearty, Hekker, Hogg, Holtzman, Horta, Hou, Hsieh, Huber,
  Hunt, Chitham, Imig, Jaber, Angel, Johnson, Jones, J{\"o}nsson, Jullo, Kim,
  Kinemuchi, Kirkpatrick, Kite, Klaene, Kneib, Kollmeier, Kong, Kounkel,
  Krishnarao, Lacerna, Lan, Lane, Law, Le~Goff, Leung, Lewis, Li, Lian, Lin,
  Long, Longa-Pe{\~n}a, Lundgren, Lyke, Ted~Mackereth, MacLeod, Majewski,
  Manchado, Maraston, Martini, Masseron, Masters, Mathur, McDermid, Merloni,
  Merrifield, M{\'e}sz{\'a}ros, Miglio, Minniti, Minsley, Miyaji, Mohammad,
  Mosser, Mueller, Muna, Mu{\~n}oz-Guti{\'e}rrez, Myers, Nadathur, Nair,
  Nandra, do~Nascimento, Nevin, Newman, Nidever, Nitschelm, Noterdaeme,
  O'Connell, Olmstead, Oravetz, Oravetz, Osorio, Pace, Padilla,
  Palanque-Delabrouille, Palicio, Pan, Pan, Parker, Paviot, Peirani,
  Ram{\'r}ez, Penny, Percival, Perez-Fournon, P{\'e}rez-R{\`a}fols, Petitjean,
  Pieri, Pinsonneault, Poovelil, Povick, Prakash, Price-Whelan, Raddick,
  Raichoor, Ray, Rembold, Rezaie, Riffel, Riffel, Rix, Robin, Roman-Lopes,
  Rom{\'a}n-Z{\'u}{\~n}iga, Rose, Ross, Rossi, Rowlands, Rubin, Salvato,
  S{\'a}nchez, S{\'a}nchez-Menguiano, S{\'a}nchez-Gallego, Sayres, Schaefer,
  Schiavon, Schimoia, Schlafly, Schlegel, Schneider, Schultheis, Schwope, Seo,
  Serenelli, Shafieloo, Shamsi, Shao, Shen, Shetrone, Shirley, Aguirre, Simon,
  Skrutskie, Slosar, Smethurst, Sobeck, Sodi, Souto, Stark, Stassun, Steinmetz,
  Stello, Stermer, Storchi-Bergmann, Streblyanska, Stringfellow, Stutz,
  Su{\'a}rez, Sun, Taghizadeh-Popp, Talbot, Tayar, Thakar, Theriault, Thomas,
  Thomas, Tinker, Tojeiro, Toledo, Tremonti, Troup, Tuttle, Unda-Sanzana,
  Valentini, Vargas-Gonz{\'a}lez, Vargas-Maga{\~n}a, V{\'a}zquez-Mata, Vivek,
  Wake, Wang, Weaver, Weijmans, Wild, Wilson, Wilson, Wolthuis, Wood-Vasey,
  Yan, Yang, Y{\`e}che, Zamora, Zarrouk, Zasowski, Zhang, Zhao, Zhao, Zheng,
  Zheng, Zhu, \& Zou}]{Ahumada2020}
Ahumada, R., Prieto, C.~A., Almeida, A., {et~al.} 2020, \apjs, 249, 3,
  \dodoi{10.3847/1538-4365/ab929e}

\bibitem[{{Badenes} \& {Maoz}(2012)}]{B+M2012}
{Badenes}, C., \& {Maoz}, D. 2012, \apjl, 749, L11,
  \dodoi{10.1088/2041-8205/749/1/L11}

\bibitem[{{Badenes} {et~al.}(2018){Badenes}, {Mazzola}, {Thompson}, {Covey},
  {Freeman}, {Walker}, {Moe}, {Troup}, {Nidever}, {Allende Prieto}, {Andrews},
  {Barb{\'a}}, {Beers}, {Bovy}, {Carlberg}, {De Lee}, {Johnson}, {Lewis},
  {Majewski}, {Pinsonneault}, {Sobeck}, {Stassun}, {Stringfellow}, \&
  {Zasowski}}]{Badenes2018}
{Badenes}, C., {Mazzola}, C., {Thompson}, T.~A., {et~al.} 2018, \apj, 854, 147.
\newblock \doarXiv{1711.00660}

\bibitem[{{Baumgardt} {et~al.}(2022){Baumgardt}, {Faller}, {Meinhold},
  {McGovern-Greco}, \& {Hilker}}]{Baumgardt22}
{Baumgardt}, H., {Faller}, J., {Meinhold}, N., {McGovern-Greco}, C., \&
  {Hilker}, M. 2022, \mnras, 510, 3531, \dodoi{10.1093/mnras/stab3629}

\bibitem[{Bentley(1975)}]{bentley1975}
Bentley, J.~L. 1975, Commun. ACM, 18, 509–517, \dodoi{10.1145/361002.361007}

\bibitem[{{Binney} \& {Tremaine}(2008)}]{2008gady.book.....B}
{Binney}, J., \& {Tremaine}, S. 2008, {Galactic Dynamics: Second Edition}

\bibitem[{Blanton {et~al.}(2017)Blanton, Bershady, Abolfathi, Albareti,
  Allende~Prieto, Almeida, Alonso-Garc{\'{\i}}a, Anders, Anderson, Andrews,
  Aquino-Ort{\'{\i}}z, Arag{\'o}n-Salamanca, Argudo-Fern{\'a}ndez, Armengaud,
  Aubourg, Avila-Reese, Badenes, Bailey, Barger, Barrera-Ballesteros, Bartosz,
  Bates, Baumgarten, Bautista, Beaton, Beers, Belfiore, Bender, Berlind,
  Bernardi, Beutler, Bird, Bizyaev, Blanc, Blomqvist, Bolton, Boquien,
  Borissova, van~den Bosch, Bovy, Brandt, Brinkmann, Brownstein, Bundy,
  Burgasser, Burtin, Busca, Cappellari, Delgado~Carigi, Carlberg,
  Carnero~Rosell, Carrera, Chanover, Cherinka, Cheung, G{\'o}mez Maqueo~Chew,
  Chiappini, Choi, Chojnowski, Chuang, Chung, Cirolini, Clerc, Cohen, Comparat,
  da~Costa, Cousinou, Covey, Crane, Croft, Cruz-Gonzalez, Garrido~Cuadra,
  Cunha, Damke, Darling, Davies, Dawson, de~la Macorra, Dell'Agli, De~Lee,
  Delubac, Di~Mille, Diamond-Stanic, Cano-D{\'{\i}}az, Donor, Downes, Drory,
  du~Mas~des Bourboux, Duckworth, Dwelly, Dyer, Ebelke, Eigenbrot, Eisenstein,
  Emsellem, Eracleous, Escoffier, Evans, Fan, Fern{\'a}ndez-Alvar,
  Fernandez-Trincado, Feuillet, Finoguenov, Fleming, Font-Ribera, Fredrickson,
  Freischlad, Frinchaboy, Fuentes, Galbany, Garcia-Dias,
  Garc{\'{\i}}a-Hern{\'a}ndez, Gaulme, Geisler, Gelfand, Gil-Mar{\'{\i}}n,
  Gillespie, Goddard, Gonzalez-Perez, Grabowski, Green, Grier, Gunn, Guo, Guy,
  Hagen, Hahn, Hall, Harding, Hasselquist, Hawley, Hearty,
  Gonzalez~Hern{\'a}ndez, Ho, Hogg, Holley-Bockelmann, Holtzman, Holzer,
  Huehnerhoff, Hutchinson, Hwang, Ibarra-Medel, da~Silva~Ilha, Ivans, Ivory,
  Jackson, Jensen, Johnson, Jones, J{\"o}nsson, Jullo, Kamble, Kinemuchi,
  Kirkby, Kitaura, Klaene, Knapp, Kneib, Kollmeier, Lacerna, Lane, Lang, Law,
  Lazarz, Lee, Le~Goff, Liang, Li, Li, Lian, Lima, Lin, Lin, Bertran~de Lis,
  Liu, de~Icaza~Lizaola, Long, Lucatello, Lundgren, MacDonald, Deconto~Machado,
  MacLeod, Mahadevan, Geimba~Maia, Maiolino, Majewski, Malanushenko,
  Malanushenko, Manchado, Mao, Maraston, Marques-Chaves, Masseron, Masters,
  McBride, McDermid, McGrath, McGreer, Medina~Pe{\~n}a, Melendez, Merloni,
  Merrifield, Meszaros, Meza, Minchev, Minniti, Miyaji, More, Mulchaey,
  M{\"u}ller-S{\'a}nchez, Muna, Munoz, Myers, Nair, Nandra, Correa~do
  Nascimento, Negrete, Ness, Newman, Nichol, Nidever, Nitschelm, Ntelis,
  O'Connell, Oelkers, Oravetz, Oravetz, Pace, Padilla, Palanque-Delabrouille,
  Alonso~Palicio, Pan, Parejko, Parikh, P{\^a}ris, Park, Patten, Peirani,
  Pellejero-Ibanez, Penny, Percival, Perez-Fournon, Petitjean, Pieri,
  Pinsonneault, Pisani, Poleski, Prada, Prakash, Queiroz, Raddick, Raichoor,
  Barboza~Rembold, Richstein, Riffel, Riffel, Rix, Robin, Rockosi,
  Rodr{\'{\i}}guez-Torres, Roman-Lopes, Rom{\'a}n-Z{\'u}{\~n}iga, Rosado, Ross,
  Rossi, Ruan, Ruggeri, Rykoff, Salazar-Albornoz, Salvato, S{\'a}nchez, Aguado,
  S{\'a}nchez-Gallego, Santana, Santiago, Sayres, Schiavon, da~Silva~Schimoia,
  Schlafly, Schlegel, Schneider, Schultheis, Schuster, Schwope, Seo, Shao,
  Shen, Shetrone, Shull, Simon, Skinner, Skrutskie, Slosar, Smith, Sobeck,
  Sobreira, Somers, Souto, Stark, Stassun, Stauffer, Steinmetz,
  Storchi-Bergmann, Streblyanska, Stringfellow, Su{\'a}rez, Sun, Suzuki,
  Szigeti, Taghizadeh-Popp, Tang, Tao, Tayar, Tembe, Teske, Thakar, Thomas,
  Thompson, Tinker, Tissera, Tojeiro, Hernandez~Toledo, de~la Torre, Tremonti,
  Troup, Valenzuela, Martinez~Valpuesta, Vargas-Gonz{\'a}lez,
  Vargas-Maga{\~n}a, Vazquez, Villanova, Vivek, Vogt, Wake, Walterbos, Wang,
  Weaver, Weijmans, Weinberg, Westfall, Whelan, Wild, Wilson, Wood-Vasey,
  Wylezalek, Xiao, Yan, Yang, Ybarra, Y{\`e}che, Zakamska, Zamora, Zarrouk,
  Zasowski, Zhang, Zhao, Zheng, Zheng, Zhou, Zhou, Zhu, Zoccali, \&
  Zou}]{Blanton2017}
Blanton, M.~R., Bershady, M.~A., Abolfathi, B., {et~al.} 2017, The Astronomical
  Journal, 154, 28, \dodoi{10.3847/1538-3881/aa7567}

\bibitem[{{Buttry} {et~al.}(2021){Buttry}, {Pace}, {Koposov}, {Walker},
  {Caldwell}, {Kirby}, {Martin}, {Mateo}, {Olszewski}, {Starkenburg},
  {Badenes}, \& {Mazzola Daher}}]{Buttry+21}
{Buttry}, R., {Pace}, A.~B., {Koposov}, S.~E., {et~al.} 2021, arXiv e-prints,
  arXiv:2108.10867.
\newblock \doarXiv{2108.10867}

\bibitem[{{Fielder} {et~al.}(2021){Fielder}, {Newman}, {Andrews}, {Zasowski},
  {Boardman}, {Licquia}, {Masters}, \& {Salim}}]{fielder2021}
{Fielder}, C.~E., {Newman}, J.~A., {Andrews}, B.~H., {et~al.} 2021, arXiv
  e-prints, arXiv:2106.14900.
\newblock \doarXiv{2106.14900}

\bibitem[{{Gaia Collaboration} {et~al.}(2018){Gaia Collaboration}, {Helmi},
  {van Leeuwen}, {McMillan}, {Massari}, {Antoja}, {Robin}, {Lindegren},
  {Bastian}, {Arenou}, {Babusiaux}, {Biermann}, {Breddels}, {Hobbs}, {Jordi},
  {Pancino}, {Reyl{\'e}}, {Veljanoski}, {Brown}, {Vallenari}, {Prusti}, {de
  Bruijne}, {Bailer-Jones}, {Evans}, {Eyer}, {Jansen}, {Klioner}, {Lammers},
  {Luri}, {Mignard}, {Panem}, {Pourbaix}, {Randich}, {Sartoretti}, {Siddiqui},
  {Soubiran}, {Walton}, {Cropper}, {Drimmel}, {Katz}, {Lattanzi}, {Bakker},
  {Cacciari}, {Casta{\~n}eda}, {Chaoul}, {Cheek}, {De Angeli}, {Fabricius},
  {Guerra}, {Holl}, {Masana}, {Messineo}, {Mowlavi}, {Nienartowicz}, {Panuzzo},
  {Portell}, {Riello}, {Seabroke}, {Tanga}, {Th{\'e}venin}, {Gracia-Abril},
  {Comoretto}, {Garcia-Reinaldos}, {Teyssier}, {Altmann}, {Andrae}, {Audard},
  {Bellas-Velidis}, {Benson}, {Berthier}, {Blomme}, {Burgess}, {Busso},
  {Carry}, {Cellino}, {Clementini}, {Clotet}, {Creevey}, {Davidson}, {De
  Ridder}, {Delchambre}, {Dell'Oro}, {Ducourant},
  {Fern{\'a}ndez-Hern{\'a}ndez}, {Fouesneau}, {Fr{\'e}mat}, {Galluccio},
  {Garc{\'\i}a-Torres}, {Gonz{\'a}lez-N{\'u}{\~n}ez}, {Gonz{\'a}lez-Vidal},
  {Gosset}, {Guy}, {Halbwachs}, {Hambly}, {Harrison}, {Hern{\'a}ndez},
  {Hestroffer}, {Hodgkin}, {Hutton}, {Jasniewicz}, {Jean-Antoine-Piccolo},
  {Jordan}, {Korn}, {Krone-Martins}, {Lanzafame}, {Lebzelter}, {L{\"o}ffler},
  {Manteiga}, {Marrese}, {Mart{\'\i}n-Fleitas}, {Moitinho}, {Mora}, {Muinonen},
  {Osinde}, {Pauwels}, {Petit}, {Recio-Blanco}, {Richards}, {Rimoldini},
  {Sarro}, {Siopis}, {Smith}, {Sozzetti}, {S{\"u}veges}, {Torra}, {van Reeven},
  {Abbas}, {Abreu Aramburu}, {Accart}, {Aerts}, {Altavilla}, {{\'A}lvarez},
  {Alvarez}, {Alves}, {Anderson}, {Andrei}, {Anglada Varela}, {Antiche},
  {Arcay}, {Astraatmadja}, {Bach}, {Baker}, {Balaguer-N{\'u}{\~n}ez}, {Balm},
  {Barache}, {Barata}, {Barbato}, {Barblan}, {Barklem}, {Barrado}, {Barros},
  {Barstow}, {Bartholom{\'e} Mu{\~n}oz}, {Bassilana}, {Becciani}, {Bellazzini},
  {Berihuete}, {Bertone}, {Bianchi}, {Bienaym{\'e}}, {Blanco-Cuaresma}, {Boch},
  {Boeche}, {Bombrun}, {Borrachero}, {Bossini}, {Bouquillon}, {Bourda},
  {Bragaglia}, {Bramante}, {Bressan}, {Brouillet}, {Br{\"u}semeister},
  {Brugaletta}, {Bucciarelli}, {Burlacu}, {Busonero}, {Butkevich}, {Buzzi},
  {Caffau}, {Cancelliere}, {Cannizzaro}, {Cantat-Gaudin}, {Carballo},
  {Carlucci}, {Carrasco}, {Casamiquela}, {Castellani}, {Castro-Ginard},
  {Charlot}, {Chemin}, {Chiavassa}, {Cocozza}, {Costigan}, {Cowell}, {Crifo},
  {Crosta}, {Crowley}, {Cuypers}, {Dafonte}, {Damerdji}, {Dapergolas}, {David},
  {David}, {de Laverny}, {De Luise}, {De March}, {de Martino}, {de Souza}, {de
  Torres}, {Debosscher}, {del Pozo}, {Delbo}, {Delgado}, {Delgado}, {Di
  Matteo}, {Diakite}, {Diener}, {Distefano}, {Dolding}, {Drazinos},
  {Dur{\'a}n}, {Edvardsson}, {Enke}, {Eriksson}, {Esquej}, {Eynard Bontemps},
  {Fabre}, {Fabrizio}, {Faigler}, {Falc{\~a}o}, {Farr{\`a}s Casas}, {Federici},
  {Fedorets}, {Fernique}, {Figueras}, {Filippi}, {Findeisen}, {Fonti},
  {Fraile}, {Fraser}, {Fr{\'e}zouls}, {Gai}, {Galleti}, {Garabato},
  {Garc{\'\i}a-Sedano}, {Garofalo}, {Garralda}, {Gavel}, {Gavras}, {Gerssen},
  {Geyer}, {Giacobbe}, {Gilmore}, {Girona}, {Giuffrida}, {Glass}, {Gomes},
  {Granvik}, {Gueguen}, {Guerrier}, {Guiraud}, {Guti{\'e}rrez-S{\'a}nchez},
  {Hofmann}, {Holland}, {Huckle}, {Hypki}, {Icardi}, {Jan{\ss}en}, {Jevardat de
  Fombelle}, {Jonker}, {Juh{\'a}sz}, {Julbe}, {Karampelas}, {Kewley}, {Klar},
  {Kochoska}, {Kohley}, {Kolenberg}, {Kontizas}, {Kontizas}, {Koposov},
  {Kordopatis}, {Kostrzewa-Rutkowska}, {Koubsky}, {Lambert}, {Lanza}, {Lasne},
  {Lavigne}, {Le Fustec}, {Le Poncin-Lafitte}, {Lebreton}, {Leccia}, {Leclerc},
  {Lecoeur-Taibi}, {Lenhardt}, {Leroux}, {Liao}, {Licata}, {Lindstr{\o}m},
  {Lister}, {Livanou}, {Lobel}, {L{\'o}pez}, {Managau}, {Mann}, {Mantelet},
  {Marchal}, {Marchant}, {Marconi}, {Marinoni}, {Marschalk{\'o}}, {Marshall},
  {Martino}, {Marton}, {Mary}, {Matijevi{\v{c}}}, {Mazeh}, {Messina},
  {Michalik}, {Millar}, {Molina}, {Molinaro}, {Moln{\'a}r}, {Montegriffo},
  {Mor}, {Morbidelli}, {Morel}, {Morris}, {Mulone}, {Muraveva}, {Musella},
  {Nelemans}, {Nicastro}, {Noval}, {O'Mullane}, {Ord{\'e}novic},
  {Ord{\'o}{\~n}ez-Blanco}, {Osborne}, {Pagani}, {Pagano}, {Pailler},
  {Palacin}, {Palaversa}, {Panahi}, {Pawlak}, {Piersimoni}, {Pineau}, {Plachy},
  {Plum}, {Poggio}, {Poujoulet}, {Pr{\v{s}}a}, {Pulone}, {Racero}, {Ragaini},
  {Rambaux}, {Ramos-Lerate}, {Regibo}, {Riclet}, {Ripepi}, {Riva}, {Rivard},
  {Rixon}, {Roegiers}, {Roelens}, {Romero-G{\'o}mez}, {Rowell}, {Royer},
  {Ruiz-Dern}, {Sadowski}, {Sagrist{\`a} Sell{\'e}s}, {Sahlmann}, {Salgado},
  {Salguero}, {Sanna}, {Santana-Ros}, {Sarasso}, {Savietto}, {Schultheis},
  {Sciacca}, {Segol}, {Segovia}, {S{\'e}gransan}, {Shih}, {Siltala}, {Silva},
  {Smart}, {Smith}, {Solano}, {Solitro}, {Sordo}, {Soria Nieto}, {Souchay},
  {Spagna}, {Spoto}, {Stampa}, {Steele}, {Steidelm{\"u}ller}, {Stephenson},
  {Stoev}, {Suess}, {Surdej}, {Szabados}, {Szegedi-Elek}, {Tapiador}, {Taris},
  {Tauran}, {Taylor}, {Teixeira}, {Terrett}, {Teyssandier}, {Thuillot},
  {Titarenko}, {Torra Clotet}, {Turon}, {Ulla}, {Utrilla}, {Uzzi}, {Vaillant},
  {Valentini}, {Valette}, {van Elteren}, {Van Hemelryck}, {van Leeuwen},
  {Vaschetto}, {Vecchiato}, {Viala}, {Vicente}, {Vogt}, {von Essen}, {Voss},
  {Votruba}, {Voutsinas}, {Walmsley}, {Weiler}, {Wertz}, {Wevems},
  {Wyrzykowski}, {Yoldas}, {{\v{Z}}erjal}, {Ziaeepour}, {Zorec}, {Zschocke},
  {Zucker}, {Zurbach}, \& {Zwitter}}]{gaia}
{Gaia Collaboration}, {Helmi}, A., {van Leeuwen}, F., {et~al.} 2018, \aap, 616,
  A12, \dodoi{10.1051/0004-6361/201832698}

\bibitem[{{Garc{\'{\i}}a P{\'{e}}rez} {et~al.}(2016){Garc{\'{\i}}a
  P{\'{e}}rez}, {Allende Prieto}, Holtzman, Shetrone, M{\'{e}}sz{\'{a}}ros,
  Bizyaev, Carrera, Cunha, Garc{\'{\i}}a-Hern{\'{a}}ndez, Johnson, Majewski,
  Nidever, Schiavon, Shane, Smith, Sobeck, Troup, Zamora, Weinberg, Bovy,
  Eisenstein, Feuillet, Frinchaboy, Hayden, Hearty, Nguyen, O'Connell,
  Pinsonneault, Wilson, \& Zasowski}]{Perez2016}
{Garc{\'{\i}}a P{\'{e}}rez}, A.~E., {Allende Prieto}, C., Holtzman, J.~A.,
  {et~al.} 2016, \aj, 151, 144, \dodoi{10.3847/0004-6256/151/6/144}

\bibitem[{{Gibbons} {et~al.}(2017){Gibbons}, {Belokurov}, \&
  {Evans}}]{Gibbons2017}
{Gibbons}, S.~L.~J., {Belokurov}, V., \& {Evans}, N.~W. 2017, \mnras, 464, 794,
  \dodoi{10.1093/mnras/stw2328}

\bibitem[{Gunn {et~al.}(2006)Gunn, Siegmund, Mannery, Owen, Hull, Leger, Carey,
  Knapp, York, Boroski, Kent, Lupton, Rockosi, Evans, Waddell, Anderson, Annis,
  Barentine, Bartoszek, Bastian, Bracker, Brewington, Briegel, Brinkmann,
  Brown, Carr, Czarapata, Drennan, Dombeck, Federwitz, Gillespie, Gonzales,
  Hansen, Harvanek, Hayes, Jordan, Kinney, Klaene, Kleinman, Kron, Kresinski,
  Lee, Limmongkol, Lindenmeyer, Long, Loomis, McGehee, Mantsch, Eric
  H.~Neilsen, Neswold, Newman, Nitta, John~Peoples, Pier, Prieto, Prosapio,
  Rivetta, Schneider, Snedden, \& i~Wang}]{Gunn2006}
Gunn, J.~E., Siegmund, W.~A., Mannery, E.~J., {et~al.} 2006, The Astronomical
  Journal, 131, 2332, \dodoi{10.1086/500975}

\bibitem[{{Hasselquist} {et~al.}(2017){Hasselquist}, {Shetrone}, {Smith},
  {Holtzman}, {McWilliam}, {Fern{\'a}ndez-Trincado}, {Beers}, {Majewski},
  {Nidever}, {Tang}, {Tissera}, {Fern{\'a}ndez Alvar}, {Allende Prieto},
  {Almeida}, {Anguiano}, {Battaglia}, {Carigi}, {Delgado Inglada},
  {Frinchaboy}, {Garc{\'\i}a-Hern{\'a}ndez}, {Geisler}, {Minniti}, {Placco},
  {Schultheis}, {Sobeck}, \& {Villanova}}]{Has}
{Hasselquist}, S., {Shetrone}, M., {Smith}, V., {et~al.} 2017, \apj, 845, 162,
  \dodoi{10.3847/1538-4357/aa7ddc}

\bibitem[{{Hayes} {et~al.}(2020){Hayes}, {Majewski}, {Hasselquist}, {Anguiano},
  {Shetrone}, {Law}, {Schiavon}, {Cunha}, {Smith}, {Beaton}, {Price-Whelan},
  {Allende Prieto}, {Battaglia}, {Bizyaev}, {Brownstein}, {Cohen},
  {Frinchaboy}, {Garc{\'i}a-Hern{\'a}ndez}, {Lacerna}, {Lane}, {Bidin},
  {M{\~u}noz}, {Nidever}, {Oravetz}, {Oravetz}, {Pan}, {Roman-Lopes}, {Sobeck},
  \& {Stringfellow}}]{Hayes}
{Hayes}, C.~R., {Majewski}, S.~R., {Hasselquist}, S., {et~al.} 2020, \apj, 889,
  \dodoi{10.3847/1538-4357/ab62ad}

\bibitem[{Holtzman {et~al.}(2015)Holtzman, Shetrone, Johnson, Prieto, Anders,
  Andrews, Beers, Bizyaev, Blanton, Bovy, Carrera, Chojnowski, Cunha,
  Eisenstein, Feuillet, Frinchaboy, Galbraith-Frew, P{\'{e}}rez,
  Garc{\'{\i}}a-Hern{\'{a}}ndez, Hasselquist, Hayden, Hearty, Ivans, Majewski,
  Martell, Meszaros, Muna, Nidever, Nguyen, O'Connell, Pan, Pinsonneault,
  Robin, Schiavon, Shane, Sobeck, Smith, Troup, Weinberg, Wilson, Wood-Vasey,
  Zamora, \& Zasowski}]{Holtzman2015}
Holtzman, J.~A., Shetrone, M., Johnson, J.~A., {et~al.} 2015, The Astronomical
  Journal, 150, 148, \dodoi{10.1088/0004-6256/150/5/148}

\bibitem[{{Hunter}(2007)}]{matplotlib}
{Hunter}, J.~D. 2007, Computing in Science Engineering, 9, 90

\bibitem[{{Ibata} {et~al.}(1994){Ibata}, {Gilmore}, \& {Irwin}}]{Ibata}
{Ibata}, R.~A., {Gilmore}, G., \& {Irwin}, M.~J. 1994, \nat, 370, 194

\bibitem[{Jönsson {et~al.}(2020)Jönsson, Holtzman, Prieto, Cunha,
  Garc{\'{\i}}a-Hern{\'{a}}ndez, Hasselquist, Masseron, Osorio, Shetrone,
  Smith, Stringfellow, Bizyaev, Edvardsson, Majewski, M{\'{e}}sz{\'{a}}ros,
  Souto, Zamora, Beaton, Bovy, Donor, Pinsonneault, Poovelil, \&
  Sobeck}]{Joensson2020}
Jönsson, H., Holtzman, J.~A., Prieto, C.~A., {et~al.} 2020, \aj, 160, 120,
  \dodoi{10.3847/1538-3881/aba592}

\bibitem[{{Koposov} {et~al.}(2011){Koposov}, {Gilmore}, {Walker}, {Belokurov},
  {Evans}, {Fellhauer}, {Gieren}, {Geisler}, {Monaco}, {Norris}, {Okamoto},
  {Pe{\~n}arrubia}, {Wilkinson}, {Wyse}, \& {Zucker}}]{koposov+11}
{Koposov}, S.~E., {Gilmore}, G., {Walker}, M.~G., {et~al.} 2011, \apj, 736,
  146, \dodoi{10.1088/0004-637X/736/2/146}

\bibitem[{Kramida {et~al.}(2021)Kramida, {Yu.~Ralchenko}, Reader, \& {and NIST
  ASD Team}}]{NIST_ASD}
Kramida, A., {Yu.~Ralchenko}, Reader, J., \& {and NIST ASD Team}. 2021, {NIST
  Atomic Spectra Database (ver. 5.9), [Online]. Available:
  {\tt{https://physics.nist.gov/asd}} [2022, January 10]. National Institute of
  Standards and Technology, Gaithersburg, MD.}

\bibitem[{{Licquia} {et~al.}(2015){Licquia}, {Newman}, \&
  {Brinchmann}}]{licquia2015}
{Licquia}, T.~C., {Newman}, J.~A., \& {Brinchmann}, J. 2015, \apj, 809, 96,
  \dodoi{10.1088/0004-637X/809/1/96}

\bibitem[{Majewski {et~al.}(2017)Majewski, Schiavon, Frinchaboy, Prieto,
  Barkhouser, Bizyaev, Blank, Brunner, Burton, Carrera, Chojnowski, Cunha,
  Epstein, Fitzgerald, P{\'{e}}rez, Hearty, Henderson, Holtzman, Johnson, Lam,
  Lawler, Maseman, M{\'{e}}sz{\'{a}}ros, Nelson, Nguyen, Nidever, Pinsonneault,
  Shetrone, Smee, Smith, Stolberg, Skrutskie, Walker, Wilson, Zasowski, Anders,
  Basu, Beland, Blanton, Bovy, Brownstein, Carlberg, Chaplin, Chiappini,
  Eisenstein, Elsworth, Feuillet, Fleming, Galbraith-Frew, Garc{\'{\i}}a,
  Garc{\'{\i}}a-Hern{\'{a}}ndez, Gillespie, Girardi, Gunn, Hasselquist, Hayden,
  Hekker, Ivans, Kinemuchi, Klaene, Mahadevan, Mathur, Mosser, Muna, Munn,
  Nichol, O'Connell, Parejko, Robin, Rocha-Pinto, Schultheis, Serenelli, Shane,
  Aguirre, Sobeck, Thompson, Troup, Weinberg, \& Zamora}]{Majewski2017}
Majewski, S.~R., Schiavon, R.~P., Frinchaboy, P.~M., {et~al.} 2017, The
  Astronomical Journal, 154, 94, \dodoi{10.3847/1538-3881/aa784d}

\bibitem[{{Maoz} {et~al.}(2012){Maoz}, {Badenes}, \& {Bickerton}}]{M+B+B2012}
{Maoz}, D., {Badenes}, C., \& {Bickerton}, S.~J. 2012, \apj, 751, 143,
  \dodoi{10.1088/0004-637X/751/2/143}

\bibitem[{{Martinez} {et~al.}(2011){Martinez}, {Minor}, {Bullock},
  {Kaplinghat}, {Simon}, \& {Geha}}]{martinez+11}
{Martinez}, G.~D., {Minor}, Q.~E., {Bullock}, J., {et~al.} 2011, \apj, 738, 55,
  \dodoi{10.1088/0004-637X/738/1/55}

\bibitem[{{Mazzola} {et~al.}(2020){Mazzola}, {Badenes}, {Moe}, {Koposov},
  {Kounkel}, {Kratter}, {Covey}, {Walker}, {Thompson}, {Andrews}, {Freeman},
  {Anguiano}, {Carlberg}, {De Lee}, {Frinchaboy}, {Lewis}, {Majewski},
  {Nidever}, {Nitschelm}, {Price-Whelan}, {Roman-Lopes}, {Stassun}, \&
  {Troup}}]{Mazzola}
{Mazzola}, C., {Badenes}, C., {Moe}, M., {et~al.} 2020, Monthly Notices of the
  Royal Astronomical Society, 499, \dodoi{10.1093/mnras/staa2859}

\bibitem[{{McConnachie} \& {C{\^o}t{\'e}}(2010)}]{mcconnahie+cote+10}
{McConnachie}, A.~W., \& {C{\^o}t{\'e}}, P. 2010, \apjl, 722, L209,
  \dodoi{10.1088/2041-8205/722/2/L209}

\bibitem[{{Minor} {et~al.}(2019){Minor}, {Pace}, {Marshall}, \&
  {Strigari}}]{minor+19}
{Minor}, Q.~E., {Pace}, A.~B., {Marshall}, J.~L., \& {Strigari}, L.~E. 2019,
  \mnras, 487, 2961, \dodoi{10.1093/mnras/stz1468}

\bibitem[{{Moe} \& {Di Stefano}(2017)}]{moe+di+2017}
{Moe}, M., \& {Di Stefano}, R. 2017, \apjs, 230, 15,
  \dodoi{10.3847/1538-4365/aa6fb6}

\bibitem[{{Moe} {et~al.}(2019){Moe}, {Kratter}, \& {Badenes}}]{Moe+19}
{Moe}, M., {Kratter}, K.~M., \& {Badenes}, C. 2019, \apj, 875, 61,
  \dodoi{10.3847/1538-4357/ab0d88}

\bibitem[{Nidever {et~al.}(2015)Nidever, Holtzman, Prieto, Beland, Bender,
  Bizyaev, Burton, Desphande, Fleming, P{\'{e}}rez, Hearty, Majewski,
  M{\'{e}}sz{\'{a}}ros, Muna, Nguyen, Schiavon, Shetrone, Skrutskie, Sobeck, \&
  Wilson}]{Nidever2015}
Nidever, D.~L., Holtzman, J.~A., Prieto, C.~A., {et~al.} 2015, The Astronomical
  Journal, 150, 173, \dodoi{10.1088/0004-6256/150/6/173}

\bibitem[{Pedregosa {et~al.}(2011)Pedregosa, Varoquaux, Gramfort, Michel,
  Thirion, Grisel, Blondel, Prettenhofer, Weiss, Dubourg, Vanderplas, Passos,
  Cournapeau, Brucher, Perrot, \& Duchesnay}]{scikit-learn}
Pedregosa, F., Varoquaux, G., Gramfort, A., {et~al.} 2011, Journal of Machine
  Learning Research, 12, 2825

\bibitem[{{Perez} \& {Granger}(2007)}]{ipython}
{Perez}, F., \& {Granger}, B.~E. 2007, Computing in Science Engineering, 9, 21

\bibitem[{{Price-Whelan} {et~al.}(2017){Price-Whelan}, {Hogg},
  {Foreman-Mackey}, \& {Rix}}]{Price-Whelan2017}
{Price-Whelan}, A.~M., {Hogg}, D.~W., {Foreman-Mackey}, D., \& {Rix}, H.-W.
  2017, \apj, 837, 20, \dodoi{10.3847/1538-4357/aa5e50}

\bibitem[{{Price-Whelan} {et~al.}(2020){Price-Whelan}, {Hogg}, {Rix}, {Beaton},
  {Lewis}, {Nidever}, {Almeida}, {Badenes}, {Barba}, {Beers}, {Carlberg}, {De
  Lee}, {Fern{\'a}ndez-Trincado}, {Frinchaboy}, {Garc{\'\i}a-Hern{\'a}ndez},
  {Green}, {Hasselquist}, {Longa-Pe{\~n}a}, {Majewski}, {Nitschelm}, {Sobeck},
  {Stassun}, {Stringfellow}, \& {Troup}}]{Price-Whelan+20}
{Price-Whelan}, A.~M., {Hogg}, D.~W., {Rix}, H.-W., {et~al.} 2020, \apj, 895

\bibitem[{{Sanders} {et~al.}(2021){Sanders}, {Belokurov}, \&
  {Man}}]{sanders+21}
{Sanders}, J.~L., {Belokurov}, V., \& {Man}, K. T.~F. 2021, \mnras, 506, 4321,
  \dodoi{10.1093/mnras/stab1951}

\bibitem[{{Sanders} \& {Das}(2018)}]{sanders+das+18}
{Sanders}, J.~L., \& {Das}, P. 2018, \mnras, 481, 4093,
  \dodoi{10.1093/mnras/sty2490}

\bibitem[{{Sbordone} {et~al.}(2007){Sbordone}, {Bonifacio}, {Buonanno},
  {Marconi}, {Monaco}, \& {Zaggia}}]{Sbordone07}
{Sbordone}, L., {Bonifacio}, P., {Buonanno}, R., {et~al.} 2007, \aap, 465, 815,
  \dodoi{10.1051/0004-6361:20066385}

\bibitem[{{Simon} {et~al.}(2011){Simon}, {Geha}, {Minor}, {Martinez}, {Kirby},
  {Bullock}, {Kaplinghat}, {Strigari}, {Willman}, {Choi}, {Tollerud}, \&
  {Wolf}}]{simon+11}
{Simon}, J.~D., {Geha}, M., {Minor}, Q.~E., {et~al.} 2011, \apj, 733, 46,
  \dodoi{10.1088/0004-637X/733/1/46}

\bibitem[{{van der Walt} {et~al.}(2011){van der Walt}, {Colbert}, \&
  {Varoquaux}}]{numpy}
{van der Walt}, S., {Colbert}, S.~C., \& {Varoquaux}, G. 2011, Computing in
  Science Engineering, 13, 22

\bibitem[{{Virtanen} {et~al.}(2020){Virtanen}, {Gommers}, {Oliphant},
  {Haberland}, {Reddy}, {Cournapeau}, {Burovski}, {Peterson}, {Weckesser},
  {Bright}, {van der Walt}, {Brett}, {Wilson}, {Jarrod Millman}, {Mayorov},
  {Nelson}, {Jones}, {Kern}, {Larson}, {Carey}, {Polat}, {Feng}, {Moore}, {Vand
  erPlas}, {Laxalde}, {Perktold}, {Cimrman}, {Henriksen}, {Quintero}, {Harris},
  {Archibald}, {Ribeiro}, {Pedregosa}, {van Mulbregt}, \&
  {Contributors}}]{SciPy2020}
{Virtanen}, P., {Gommers}, R., {Oliphant}, T.~E., {et~al.} 2020, Nature
  Methods, \dodoi{https://doi.org/10.1038/s41592-019-0686-2}

\bibitem[{{W}es {M}c{K}inney(2010)}]{pandas}
{W}es {M}c{K}inney. 2010, in {P}roceedings of the 9th {P}ython in {S}cience
  {C}onference, ed. {S}t\'efan van~der {W}alt \& {J}arrod {M}illman, 56 -- 61,
  \dodoi{10.25080/Majora-92bf1922-00a}

\bibitem[{Wilson {et~al.}(2019)Wilson, Hearty, Skrutskie, Majewski, Holtzman,
  Eisenstein, Gunn, Blank, Henderson, Smee, Nelson, Nidever, Arns, Barkhouser,
  Barr, Beland, Bershady, Blanton, Brunner, Burton, Carey, Carr, Colque, Crane,
  Damke, Davidson, Dean, Mille, Don, Ebelke, Evans, Fitzgerald, Gillespie,
  Hall, Harding, Harding, Hammond, Hancock, Harrison, Hope, Horne, Karakla,
  Lam, Leger, MacDonald, Maseman, Matsunari, Melton, Mitcheltree, O'Brien,
  O'Connell, Patten, Richardson, Rieke, Rieke, Roman-Lopes, Schiavon, Sobeck,
  Stolberg, Stoll, Tembe, Trujillo, Uomoto, Vernieri, Walker, Weinberg, Young,
  Anthony-Brumfield, Bizyaev, Breslauer, Lee, Downey, Halverson, Huehnerhoff,
  Klaene, Leon, Long, Mahadevan, Malanushenko, Nguyen, Owen,
  S{\'{a}}nchez-Gallego, Sayres, Shane, Shectman, Shetrone, Skinner, Stauffer,
  \& Zhao}]{Wilson2019}
Wilson, J.~C., Hearty, F.~R., Skrutskie, M.~F., {et~al.} 2019, Publications of
  the Astronomical Society of the Pacific, 131, 055001,
  \dodoi{10.1088/1538-3873/ab0075}

\bibitem[{{Wu} \& {Zhao}(2021)}]{Wu2021}
{Wu}, W., \& {Zhao}, G. 2021, \apj, 908, 224, \dodoi{10.3847/1538-4357/abd6b8}

\bibitem[{{Wyse} {et~al.}(2020){Wyse}, {Moe}, \& {Kratter}}]{wyse+20}
{Wyse}, R. F.~G., {Moe}, M., \& {Kratter}, K.~M. 2020, \mnras, 493, 6109,
  \dodoi{10.1093/mnras/staa731}

\end{thebibliography}
\bibliographystyle{aasjournal}

\end{document}